\documentclass[runningheads]{llncs}
\usepackage{graphicx}
\usepackage{amsmath}
\usepackage{enumitem}
\usepackage{microtype}
\usepackage{hyperref} 
\usepackage{booktabs}
\usepackage{multirow}
\usepackage{soul}
\usepackage[ruled]{algorithm2e}
\usepackage{xcolor} 
\usepackage{floatrow}
\usepackage{array}
\newcolumntype{H}{>{\setbox0=\hbox\bgroup}c<{\egroup}@{}}

\begin{document}
\title{3D-UCaps: 3D Capsules Unet \\ for Volumetric Image Segmentation}
\author{Tan Nguyen\inst{1} \and
Binh-Son Hua\inst{1} \and
Ngan Le\inst{2}}

%index{Nguyen, Tan}
%index{Hua, Binh-Son}
%index{Le, Ngan}

\authorrunning{T. Nguyen et al.}
\institute{VinAI Research, Vietnam \\ \email{\{v.tannh10,v.sonhb\}@vinai.io} \and
Department of Computer Science and Computer Engineering, \\ University of Arkansas, Fayetteville USA 72701\\
\email{thile@uark.edu}}
\maketitle              % typeset the header of the contribution
\begin{abstract}
Medical image segmentation has been so far achieving promising results with Convolutional Neural Networks (CNNs).
However, it is arguable that in traditional CNNs, its pooling layer tends to discard important information such as positions. 
Moreover, CNNs are sensitive to rotation and affine transformation.
Capsule network is a data-efficient network design proposed to overcome such limitations by replacing pooling layers with dynamic routing and convolutional strides, which aims to preserve the part-whole relationships. 
Capsule network has shown a great performance in image recognition and natural language processing, but applications for medical image segmentation, particularly volumetric image segmentation, has been limited.  
In this work, we propose 3D-UCaps, a 3D voxel-based Capsule network for medical volumetric image segmentation. 
We build the concept of capsules into a CNN by designing a network with two pathways: the first pathway is encoded by 3D Capsule blocks, whereas the second pathway is decoded by 3D CNNs blocks. 
3D-UCaps, therefore inherits the merits from both Capsule network to preserve the spatial relationship and CNNs to learn visual representation.
We conducted experiments on various datasets to demonstrate the robustness of 3D-UCaps including iSeg-2017, LUNA16, Hippocampus, and Cardiac, where our method outperforms previous Capsule networks and 3D-Unets.
%, especially with volume data under rotation transformations.
Our code is available at \url{https://github.com/VinAIResearch/3D-UCaps}.

\end{abstract}

\section{Introduction}
Medical image segmentation (MIS) is a visual task that aims to identify the pixels of organs or lesions from background medical images. 
It plays a key role in medical analysis, computer-aided diagnosis, and smart medicine due to the great improvement in diagnostic efficiency and accuracy. 
Thanks to recent advances of deep learning, convolutional neural networks (CNNs) can be used to extract hierarchical feature representation for segmentation, which is robust to image degradation such as noise, blur, contrast, etc.
Among many CNNs-based segmentation approaches, FCN~\cite{long2015fully}, Unet~\cite{cciccek20163d}, and Auto-encoder-like architecture have become the desired models for MIS.
Particularly, such methods achieved impressive performance in brain tumor~\cite{brain_le_2018,brain_Le_2021_1}, liver tumor~\cite{bilic2019liver,li2018h}, optic disc~\cite{ramani2020improved,veena2020review}, retina \cite{brain_Le_2021_2},  lung~\cite{souza2019automatic,jin2018ct}, and cell~\cite{hatipoglu2017cell,moshkov2020test}. 
However, CNNs are limited in their mechanism of aggregating data at pooling layers. 
Notably, pooling summarizes features in a local window and discards important information such as pose and object location. 
Therefore, CNNs with consecutive pooling layers are unable to perverse the spatial dependencies between objects parts and wholes. 
Moreover, the activation layer plays an important role in CNNs; however, it is not interpretable and has often been used as a black box. 
MIS with CNNs is thus prone to performance degradation when data undergoes some transformations such as rotations. 
A practical example is during an MRI scan, subject motion causes transformations to appear in a subset of slices, which is a hard case for CNNs~\cite{wang2019benchmark}.

To overcome such limitations by CNNs, Sabour et al.    \cite{sabour2017dynamic}  developed a novel network architecture called Capsule Network (CapsNet). The basic idea of CapsNet is to encode the part-whole relationships (e.g., scale, locations, orientations, brightnesses) between various entities, i.e., objects, parts of objects, to achieve viewpoint equivariance. 
Unlike CNNs which learn all part features of the objects, CapsNet learns the relationship between these features through dynamically calculated weights in each forward pass. 
This optimization mechanism, i.e., dynamic routing, allows weighting the contributions of parts to a whole object differently at both training and inference. 
CapsNet has been mainly applied to image recognition; its performance is still limited compared to the state-of-the-art by CNNs-based approaches. Adapting CapsNet for semantic segmentation, e.g., SegCaps~\cite{lalonde2018capsules,lalonde2021capsules}, receives even less attention. 
In this work, we propose an effective 3D Capsules network for volumetric image segmentation, named 3D-UCaps. 
Our 3D-UCaps is built on both 3D Capsule blocks, which take temporal relations between volumetric slices into consideration, and 3D CNNs blocks, which extract contextual visual representation. 
Our 3D-UCaps contains two pathways, i.e., encoder and decoder. Whereas encoder is built upon 3D Capsule blocks, the decoder is built upon 3D CNNs blocks. 
We argue and show empirically that using deconvolutional Capsules in the decoder pathway not only reduces segmentation accuracy but also increases model complexity. 

In summary, our contributions are: 
(1) An effective 3D Capsules network for volumetric image segmentation. Our 3D-UCaps inherits the merits from both 3D Capsule block to preserve spatial relationship and 3D CNNs block to learn better visual representation.
(2) Extensive experiments on various datasets and ablation studies that showcase the effectiveness and robustness of 3D-UCaps for MIS.

\section{Background}
In CNNs, each filter of convolutional layers works like a feature detector in a small region of the input features and as we go deeper in a network, the detected low-level features are aggregated and become high-level features that can be used to distinguish between different objects. However, by doing so, each feature map only contains information about the presence of the feature, and the network relies on fixed learned weight matrix to link features between layers. It leads to the problem that the model cannot generalize well to unseen changes in the input image and usually perform poorly in that case.

CapsNet~\cite{sabour2017dynamic} is a new concept that strengthens feature learning by retaining more information at aggregation layer for pose reasoning and learning the part-whole relationship, which makes it a potential solution for semantic segmentation and object detection tasks.
Each layer in CapsNet aims to learn a set of entities (i.e., parts or objects) with their various properties and represent them in a high-dimensional form, particularly vector in \cite{sabour2017dynamic}. The length of this vector indicates the presence of the entity in the input while its orientation encodes different properties of that entity. An important assumption in CapsNet is the entity in previous layer are simple objects and based on an agreement in their votes, complex objects in next layer will be activated or not. This setting helps CapsNet reflect the changes in input through the activation of properties in the entity and still recognize the object successfully based on a dynamic voting between layers. 
Let $\{c^l_1, c^l_2, \,\ldots, c^l_n\}$ be the set of capsules in layer $l$, $\{c^{l+1}_1, c^{l+1}_2, \,\ldots, c^{l+1}_m\}$ be the set of capsule in layer $l+1$, the overall procedure will be:
\begin{align}
c^{l+1}_j = \mathrm{squash}\left( \sum_i r_{ij}v_{j|i} \right), \qquad v_{j|i} = W_{ij} c^l_i
\end{align}
where $W_{ij}$ is the learned weight matrix to linear mapping features of capsule $c^l_i$ in layer $l$ to feature space of capsule $c^{l+1}_j$ in layer $l+1$. The $r_{ij}$ are coupling coefficients between capsule $i$ and $j$ that are dynamically assigned by a routing algorithm in each forward pass such that $\sum_j r_{ij} = 1$.

SegCaps~\cite{lalonde2018capsules,lalonde2021capsules}, a state-of-the-art Capsule-based image segmentation, has made a great improvement to expand the use of CapsNet to the task of object segmentation.
This method functions by treating an MRI image as a collection of slices, each of which is then encoded and decoded by capsules to output the segmentation. 
However, SegCaps is mainly designed for 2D still images, and it performs poorly when being applied to volumetric data because of missing temporal information. 
Our work differs in that we build the CapsNet to consume 3D data directly so that both spatial and temporal information can be fully used for learning. Furthermore, our 3D-UCaps is able to take both advantages of CapsNet and 3D CNNs into consideration.

\section{Our Proposed 3D-UCaps Network}
\label{sec:method}
\begin{figure}[t]
    \centering
    \includegraphics[width=0.80\textwidth]{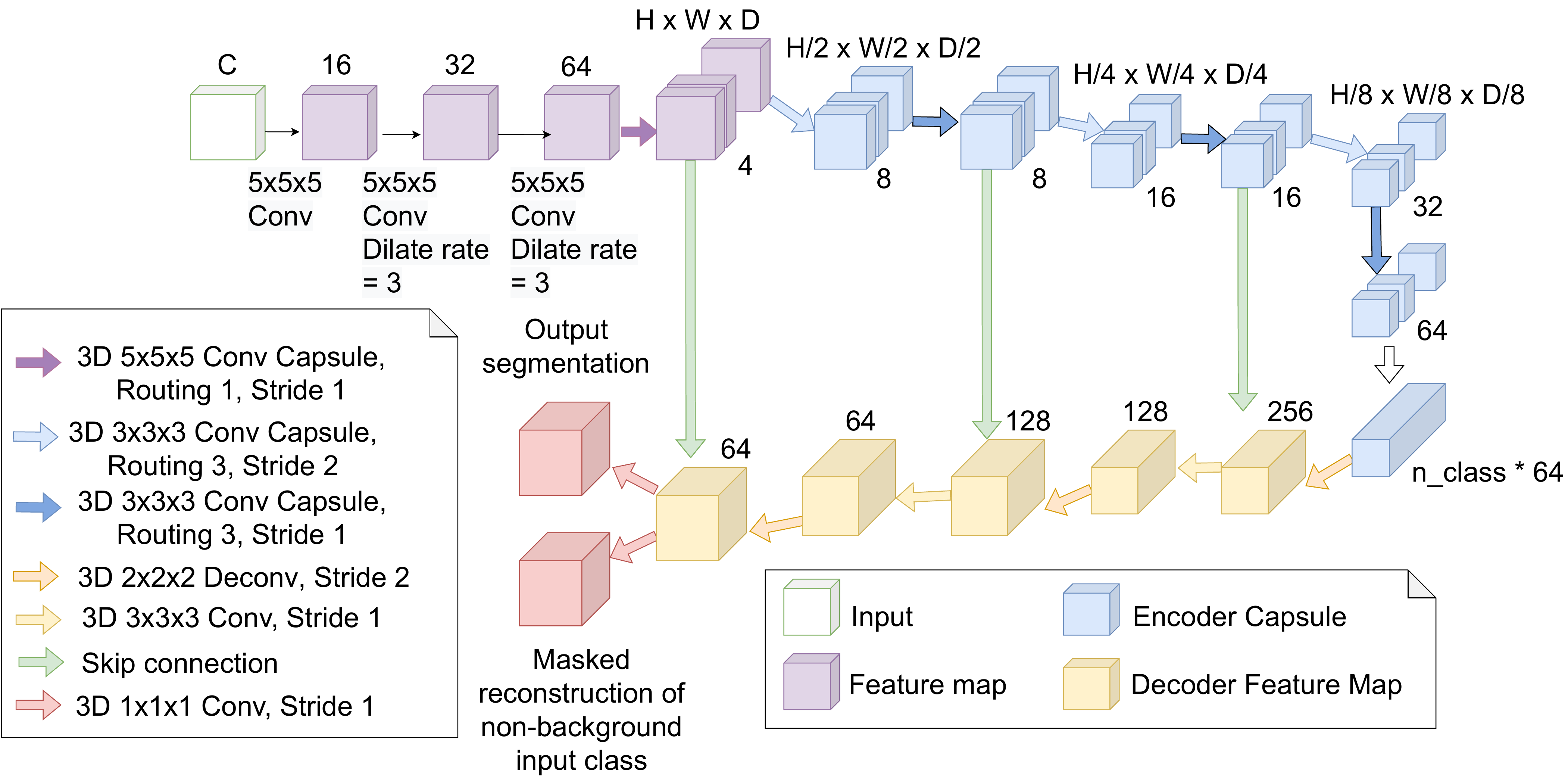}
    \caption{Our proposed 3D-UCaps architecture with three components: visual feature extraction; capsule encoder, and convolution decoder. Number on the blocks indicates number of channels in convolution layer and dimension of capsules in capsule layers. }
    \label{fig:network}
\end{figure}

In this work, we propose a hybrid 3D-UCaps network, which inherits the merits from both CapsNet and 3D CNNs. Our proposed 3D-UCaps follows Unet-like architecture \cite{cciccek20163d} and contains three main components as follows.

\noindent\textbf{Visual Feature Extractor:} We use a set of dilated convolutional layers to convert the input to high-dimensional features that can be further processed by capsules. It contains three convolution layers with the number of channels increased from 16 to 32 then 64, kernel size $5 \times 5 \times 5$ and dilate rate set to 1, 3, and 3, respectively. 
The output of this part is a feature map of size $H \times W \times D \times 64$.

\noindent\textbf{Capsule Encoder:} The visual feature from the previous component can be cast (reshaped) into a grid of $H \times W \times D$ capsules, each represented as a single 64-dimensional vector. 
Our capsule layer is a 3D convolutional capsule, which consumes volumetric data directly instead of treating it as separate slices as in SegCaps~\cite{lalonde2018capsules}. 
The advantage of our 3D capsule layer is that contextual information along the temporal axis can be included in the feature extraction.  
In additional to increasing the dimensionality of capsules as we ascend the hierarchy~\cite{sabour2017dynamic}, we suggest to use more capsule types in low-level layers and less capsule types in high-level layers. 
This is due to the fact that low-level layers represent simple object while high-level layers represent complex object and the clustering nature of routing algorithm \cite{hinton2018matrix}. 
The number of capsule types in the encoder path of our network are set to $(16, 16, 16, 8, 8, 8)$, respectively. 
This is in contrast to the design in SegCaps, where the numbers of capsules are increasing $(1, 2, 4, 4, 8, 8)$ along the encoder path.
We make sure that the number of capsule types in the last convolutional capsule layer is equal to the number of categories in the segmentation, which can be further supervised by a margin loss~\cite{sabour2017dynamic}. 
The output from a convolution capsule layer has the shape $H \times W \times D \times C \times A$, where $C$ is the number of capsule types and $A$ is the dimension of each capsule. 

\noindent\textbf{Convolutional Decoder:} We use the decoder of 3D Unet \cite{cciccek20163d} which includes deconvolution, skip connection, convolution and BatchNorm layers \cite{ioffe2015batch} to generate the segmentation from features learned by capsule layers. Particularly, we reshape the features to $H \times W \times D \times (C \star A)$ before passing them to the next convolution layer or concatenating with skip connections. 
The overall architecture can be seen in Fig. \ref{fig:network}.
Note that in our design, we only use capsule layers in the contracting path but not expanding path in the network. 
Sabour et al.~\cite{sabour2017dynamic} point out that "routing-by-agreement" should be far more effective than max-pooling, and max-pooling layers only exist in the contracting path of Unet. 

This contradicts to the design by LaLonde et al.~\cite{lalonde2018capsules}, where capsules are used in the expanding path in the network as well. 
We empirically show that using capsules in the expanding path has negligible effects compared to the traditional design while incurring high computational cost due to routing between capsule layers.

\noindent\textbf{Training Procedure.} 
We supervise our network with ground truth segmentation as follows.  
The margin loss is applied at the capsule encoder with downsampled ground truth segmentation. 
The weighted cross entropy loss is applied at the decoder head to optimize the entire network.  
To regularize the training, we also use an additional branch to output the reconstruction of the original input image as in previous work \cite{sabour2017dynamic,lalonde2018capsules}. 
We use masked mean-squared error for the reconstruction. 
The total loss is the weighted sum of the three losses. 

\section{Experimental Results}

\noindent
\textbf{Evaluation Setup}

We perform experiments on various MIS datasets to validate our method. 
Specifically, we experiment with iSeg-2017~\cite{wang2019benchmark}, LUNA16~\cite{luna}, Hippocampus, and Cardiac~\cite{simpson2019large}.
iSeg is a MRI dataset of infant brains that requires to be segmented into white matter (WM), gray matter (GM), and cerebrospinal fluid (CSF).
A recent analysis \cite{wang2019benchmark} shows that previous methods tend to perform poorly on subjects with movement and unusual poses. 
We follow the experiment setup by 3D-SkipDenseSeg \cite{bui2019skip} to conduct the report on this dataset where 9 subjects are used for training and 1 subject (subject \#9) is for testing.

Additionally, we experiment on LUNA16, Hippocampus, and Cardiac~\cite{simpson2019large} to compare with other capsule-based networks \cite{lalonde2018capsules,survarachakan2020capsule}. 
We follow a similar experiment setup in SegCaps~\cite{lalonde2018capsules} to conduct the results on LUNA16.
We also use 4-fold cross validation on training set to conduct the experiments on Hippocampus and Cardiac. 
 
 \noindent
\textbf{Implementation Details}

We implemented both 3D-SegCaps and 3D-UCaps in Pytorch. The input volumes are normalized to [0, 1]. We used patch size set as $64 \times 64 \times 64$ for iSeg and Hippocampus whereas patch size set as $128 \times 128 \times 128$ on LUNA16 and Cardiac. Both 3D-SegCaps and 3D-UCaps networks were trained without any data augmentation methods. 
We used Adam optimization with an initial learning rate of 0.0001. 
The learning rate is decayed by a factor of 0.05 if Dice score on validation set not increased for 50,000 iterations and early stopping is performed with a patience of 250,000 iterations as in \cite{lalonde2018capsules}. 
Our models were trained on NVIDIA Tesla V100 with 32GB RAM, and it takes from 2-4 days depends on the size of the dataset.

\noindent
\textbf{Performance and Comparison}

In this section, we compare our 3D-UCaps with both SOTA 3D CNNs-based segmentation approaches and other existing SegCaps methods. Furthermore, we have implemented 3D-SegCaps which is an extension version of 2D-SegCaps~\cite{lalonde2018capsules} on volumetric data to prove the effectiveness of incorporating deconvolution layers into 3D-UCaps. Our 3D-SegCaps share similar network architecture with 2D-SegCaps~\cite{lalonde2018capsules} and implemented with 3D convolution layers. This section is structured as follows: We first provide a detailed analysis on iSeg with different criteria such as segmentation accuracy, network configurations, motion artifact, and rotation invariance capability. We then report segmentation accuracy on various datasets, including LUNA16, Hippocampus, and Cardiac.

\setlength{\tabcolsep}{1.0em}
\begin{table}[t]
\centering
\caption{Comparison on iSeg-2017 dataset. The first group is 3D CNN-based networks. The second group is Capsule-based networks. The best performance is in \textbf{bold}.}
\label{table:iseg}
\begin{tabular}{@{}ll H llll@{}}
\toprule
\multicolumn{1}{c}{\multirow{2}{*}{Method}} & Depth & Params & \multicolumn{4}{c}{Dice Score}                                      \\ \cmidrule(l){4-7} 
\multicolumn{1}{c}{}                        &       &        & \multicolumn{1}{c}{WM}             & \multicolumn{1}{c}{GM}             & \multicolumn{1}{c}{CSF} & Average        \\ \midrule
Qamar et al. \cite{qamar2020variant}          & 82    & 1.7M   & 90.50                              & \textbf{92.05}                     & \textbf{95.80}          & \textbf{92.77} \\
3D-SkipDenseSeg \cite{bui2019skip}            & 47    & 1.55M  & \textbf{91.02} & 91.64 & 94.88          & 92.51 \\
VoxResNet \cite{chen2018voxresnet}                                   & 25    & 1.33M  & 89.87                              & 90.64                              & 94.28                   & 91.60          \\
3D-Unet \cite{cciccek20163d}                                     & 18    & 19M    & 89.83                              & 90.55                              & 94.39                   & 91.59          \\
CC-3D-FCN \cite{nie20183}                                   & 34    & 4.34M  & 89.19          & 90.74          & 92.40                   & 90.79          \\
DenseVoxNet \cite{jegou2017one}                                 & 32    & 4.34M  & 85.46          & 88.51          & 91.26                   & 89.24          \\ \midrule
SegCaps (2D) \cite{lalonde2018capsules}                                  & 16    & 1.4M   & 82.80                              & 84.19                              & 90.19                   & 85.73          \\
Our 3D-SegCaps                        & 16    & 6.9M   & 86.49                              & 88.53                              & 93.62                   & 89.55          \\
Our 3D-UCaps                       & 17    & 3.77M  & \textbf{90.95 }                    & \textbf{91.34}                              & \textbf{94.21 }                 & \textbf{92.17}          \\ \bottomrule
\end{tabular}
\end{table}

\underline{Accuracy:} The comparison between our proposed 3D-SegCaps, 3D-UCaps with SOTA segmentation approaches on iSeg dataset \cite{wang2019benchmark} is given in Table~\ref{table:iseg}. Thanks to taking both spatial and temporal into account, both 3D-SegCaps, 3D-UCaps outperforms 2D-SegCaps with large margin on iSeg dataset. Moreover, our 3D-UCaps consisting of Capsule encoder and Deconvolution decoder obtains better results than 3D-SegCaps, which contains both Capsule encoder and Capsule decoder. Compare to SOTA 3D CNNs networks our 3D-UCaps achieves compatible performance while our network is much shallower i.e our 3D-UCaps contains only 17 layers compare to 82 layers in \cite{qamar2020variant}. Compare to SOTA 3D CNNs networks which has similar depth, i.e. our 3D-UCaps with 18 layers, our 3D-UCaps obtains higher Dice score at individual class and on average.

\setlength{\tabcolsep}{0.8em}
\begin{table}[]
\footnotesize
\caption{Performance of 3D-UCaps on iSeg with different network configurations}
\label{table:ablation}
\centering
\begin{tabular}{lllll}
\toprule
\multicolumn{1}{c}{\multirow{2}{*}{Method}} & \multicolumn{4}{c}{Dice Score}                                                            \\ \cline{2-5} 
\multicolumn{1}{c}{}                        & \multicolumn{1}{c}{WM}    & \multicolumn{1}{c}{GM}    & \multicolumn{1}{c}{CSF} & Average \\ \hline
change number of capsule (set to 4)    & \multicolumn{1}{c}{89.02} & \multicolumn{1}{c}{89.78} & 89.95                   & 89.58   \\
without feature extractor            & 89.15                     & 89.66                     & 90.82                   & 89.88   \\
without margin loss                  & 87.62                     & 88.85                     & 92.06                   & 89.51   \\
without reconstruction loss          & 88.50                     & 88.96                     & 90.18                   & 89.22   \\
3D-UCaps                           & \textbf{90.95}            & \textbf{91.34}                     & \textbf{94.21}                   & \textbf{92.17}   \\ \bottomrule
\end{tabular}
\end{table}

\underline{Network configuration:}
To prove the effectiveness of the entire network architecture, we trained 3D-UCaps under various settings. 
The results are given in Table~\ref{table:ablation}. 
We provide a baseline where we change the number of capsules at the first layer from 16 capsules (our setting in Section~\ref{sec:method}) to 4 capsules (similar to SegCaps). 
We also examine the contribution of each component by removing feature extraction layer, margin loss, reconstruction loss, respectively. 
The result shows that each change results in accuracy drop, which validates the competence of our network model.

\setlength{\tabcolsep}{0.4em}
\begin{table}[]
\footnotesize
\caption{Performance on iSeg with motion artifact on different axis. The experiment was conducted 5 times and report average number to minimize the effect of randomization}
\label{table:slice_rotation}
\centering
\resizebox{\columnwidth}{!}{
\begin{tabular}{@{}llllllllll@{}}
\toprule
\multicolumn{1}{c}{Method} & \multicolumn{3}{c}{x-axis} & \multicolumn{3}{c}{y-axis} & \multicolumn{3}{c}{z-axis} \\ \cmidrule(l){2-10} 
                           & CSF     & GM      & WM     & CSF     & GM      & WM     & CSF     & GM      & WM     \\ \midrule
3D-SkipDenseSeg~\cite{bui2019skip}           & 83.93   & 88.76   & 88.52  & 78.98   & 87.80   & 87.89  & 82.88   & 88.38   & 88.27  \\
SegCaps (2D)~\cite{lalonde2018capsules}                & 88.11   & 83.01   & 82.01  & 86.43   & 81.80   & 80.91  & 89.36   & 83.99   & 82.76  \\
Our 3D-SegsCaps               & 90.70   & 86.15   & 84.24  & 87.75   & 84.21   & 82.76  & 89.77   & 85.54   & 83.92  \\
Our 3D-UCaps                       & \textbf{91.04}   & \textbf{88.87}   & \textbf{88.62}  & \textbf{90.31}   & \textbf{88.21}   & \textbf{88.12}  & \textbf{90.86}   & \textbf{88.65}   & \textbf{88.55}  \\ \bottomrule
\end{tabular}
}
\end{table}

\underline{Moving artifact:} Motion artifact caused by patient moving when scanning was reported as a hard case in \cite{wang2019benchmark}. We examine the influence of motion artifact to our 3D-UCaps in Table~\ref{table:slice_rotation}. In this table, motion artifact at each axis was simulated by randomly rotating 20\% number of slices along the axis with an angle between -5 and 5 degree. As can be seen, 3D-based capsules (3D-SegCaps and 3D-UCaps) both outperforms SegCaps in all classes in all rotations.

\begin{figure}[h!]
    \centering
    \includegraphics[width=\textwidth]{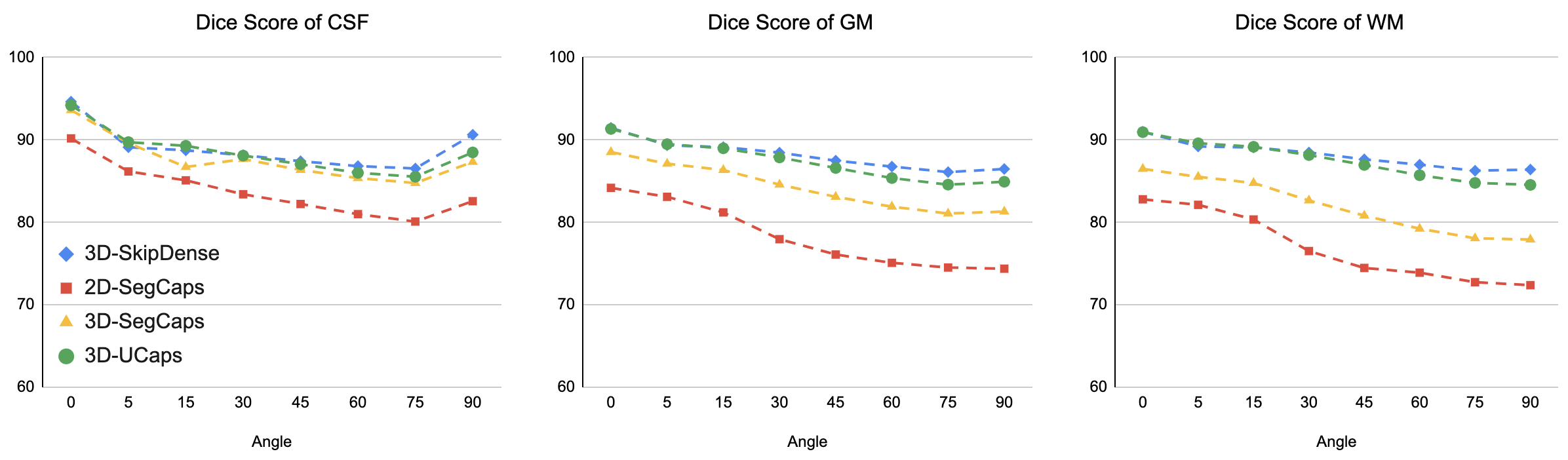}
    \caption{Comparison on iSeg with test object rotated about z-axis from zero to 90 degree. Best view in zoom.}
    \label{fig:rotation}
\end{figure}

\underline{Rotation invariance:} 
To further study rotation equivariance and invariance properties in our 3D-UCaps,
we trained our network without any rotation data augmentation. 
During testing, we choose an axis to rotate the volume, and apply the rotation with angle values fixed to 5, 10, 15, .., 90 degrees. 
Here we conduct the experiment on iSeg and choose z-axis as the rotation axis. 
We choose 3D SkipDense~\cite{bui2019skip} as 3D CNNs-based segmentation method and compare robustness to rotations between our 3D-UCaps, our 3D-SegCaps, 2D-SegCaps, and 3D SkipDense~\cite{bui2019skip}. 
The segmentation accuracy of rotation transformation on each target class is reported in Figure~\ref{fig:rotation}. 
We found that the performance tends to drop slightly when the rotation angles increases.
Except 2D-SegCaps, there is no significant difference in performance between 3D CNNs-based network and Capsule-based networks even though traditional 3D CNN-based network is not equipped with learning rotation invariance property. 
This could be explained by that the networks perform segmentation on a local patch of the volume at a time, making them resistant to local changes.
Further analysis of the robustness of capsule network on the segmentation task would be necessary, following some recent analysis on the classification task~\cite{gu2020improving,gu2021capsule}.

\underline{Results on other datasets:}
Besides iSeg, we continue benchmarking our 3D-UCaps on other datasets. The performance of 3D-UCaps on LUNA16, Hippocampus, and Cardiac is reported in Table~\ref{table:luna16}, \ref{table:hippocampus},  \ref{table:cardiac}. Different from other datasets, LUNA16 was annotated by an automated algorithm instead of a radiologist. When conducting the report on LUNA16, SegCaps \cite{lalonde2018capsules} removed 10 scans with  exceedingly poor annotations. In Table~\ref{table:luna16}, we compare our performance in two cases: full dataset and remove 10 exceedingly poor annotations.
The results show that our 3D-UCaps outperforms previous methods and our 3D-SegCaps baseline, respectively.

\setlength{\tabcolsep}{1.5em}
\begin{table}[]
\footnotesize
\caption{Comparison on LUNA16 in two cases where * indicates full dataset. The best score is in \textbf{bold}.}
\label{table:luna16}
\resizebox{\columnwidth}{!}{
\begin{tabular}{@{}l l l l l l@{}}
\toprule
Method  & Split-0 & Split-1 & Split-2 & Split-3 & Average \\ \midrule
SegCaps (2D) \cite{lalonde2018capsules} & \textbf{98.50}  & 98.52  & 98.46  & 98.47  & 98.48  \\
Our 3D-UCaps   & 98.49 & \textbf{98.61} & \textbf{98.72} & \textbf{98.76 }& \textbf{98.65} \\ \hline
SegCaps* (2D) \cite{survarachakan2020capsule} & 98.47 & 98.19 & 98.07 & 98.24 & 98.24\\
Our 3D-UCaps*   &\textbf{98.48} & \textbf{98.60}  & \textbf{98.70}  & \textbf{98.76 } & \textbf{98.64}  \\ \bottomrule
\end{tabular}
}
\end{table}

\setlength{\tabcolsep}{0.5em}
\begin{table}[]
\footnotesize
\caption{Comparison on Hippocampus dataset with 4-fold cross validation.}
\label{table:hippocampus}
\resizebox{\columnwidth}{!}{
\begin{tabular}{@{}llll|lll@{}}
\toprule
\multicolumn{1}{c}{Method} & Anterior &           &        & Posterior &           &        \\ \cmidrule(l){2-7} 
\multicolumn{1}{c}{}       & Recall   & Precision & Dice   & Recall    & Precision & Dice   \\ \midrule
%Unet (2.5D)\cite{}                       & 84.546   & 85.818    & 85.177 & 81.433    & 86.541    & 83.909 \\
Multi-SegCaps (2D)~\cite{lalonde2018capsules}               & 80.76   & 65.65    & 72.42 & 84.46    & 60.49    & 70.49 \\
EM-SegCaps (2D)~\cite{survarachakan2020capsule}                  & 17.51   & 20.01    & 18.67 & 19.00    & 34.55    & 24.52 \\
Our 3D-SegCaps                  & 94.70   & 75.41    & 83.64 & 93.09    & 73.20    & 81.67 \\
Our 3D-UCaps                       & \textbf{94.88}   & \textbf{77.48}    & \textbf{85.07} & \textbf{93.59}    & \textbf{74.03}    & \textbf{82.49} \\ \bottomrule
\end{tabular}
}
\end{table}

\setlength{\tabcolsep}{0.5em}
\begin{table}[h!]
\footnotesize
\caption{Comparison on Cardiac dataset with 4-fold cross validation.}
\label{table:cardiac}
\centering
\begin{tabular}{l l l l}
\toprule
\multicolumn{1}{c}{Method} & Recall & Precision & Dice   \\ \midrule
%Unet (2.5D)\cite{}                       & 90.690 & \underline{\textbf{92.114}}    & \underline{\textbf{91.396}} \\
SegCaps (2D)~\cite{lalonde2018capsules}                    & \textbf{96.35} & 43.96    & 60.38 \\
Multi-SegCaps (2D)~\cite{survarachakan2020capsule}              & 86.89 & 54.47    & 66.96 \\
Our 3D-SegCaps              & 88.35 & 56.40    & 67.20 \\
Our 3D-UCaps                       & 92.69 & \textbf{89.45}    & \textbf{90.82} \\ \bottomrule
\end{tabular}
\end{table}

\section{Conclusion}

In this work, we proposed a novel network architecture that can both utilize 3D capsules for learning features for volumetric segmentation while retaining the advantage of traditional convolutions in decoding the segmentation results. Even though we use capsules with dynamic routing~\cite{sabour2017dynamic,lalonde2018capsules} only in the encoder of a simple Unet like architecture, we can achieve competitive result with the state-of-the-art models on iSeg-2017 challenge while outperforming SegCaps~\cite{lalonde2018capsules} on different complex datasets. 
Exploring hybrid architecture between Capsule-based and traditional neural network is therefore a promising approach to medical image analysis while keeping model complexity and computation cost plausible. 

\vspace{3mm}
\noindent
\textbf{Acknowledgment:} 
This material is based upon work supported by the National Science Foundation under Award No. OIA-1946391.

\vspace{3mm}
\noindent
\textbf{Disclaimer:}
Any opinions, findings, and conclusions or recommendations expressed in this material are those of the author(s) and do not necessarily reflect the views of the National Science Foundation.

%\newpage
\bibliographystyle{splncs04}
\bibliography{references}

\end{document}

\title{Supplementary Material for \\3D-UCaps: 3D Capsules Unet \\ for Volumetric Image Segmentation}

\author{Tan Nguyen\inst{1} \and
Binh-Son Hua\inst{1} \and
Ngan Le\inst{2}}

%index{Nguyen, Tan}
%index{Hua, Binh-Son}
%index{Le, Ngan}

\authorrunning{T. Nguyen et al.}

\institute{VinAI Research, Vietnam \\ \email{\{v.tannh10,v.sonhb\}@vinai.io} \and
Department of Computer Science at University of Arkansas in Fayetteville \\
\email{thile@uark.edu}}
\maketitle              % typeset the header of the contribution
\begin{abstract}

\end{abstract}

Figure ~\ref{fig:visual} illustrates volumetric segmentation at three planes (sagittal, coronal, axial) on different methods given an input. The proposed Point-Unet (in the last column) provided a better segmentation results, resulting in better Dice score and Hausdorff95 distance. As can be seen, our Point-Unet segmentation provides an improved segmentation along the tumor boundary (indicated by the pink arrows) than the existing SOTA methods.

\begin{figure*}[h!]
\centering
\caption{Statistical information of Brats2018 dataset and its visualization of one subject with different modalities ($1^{st} \text{column: } T_1$, $2^{nd} \text{column: } T_{ce}$, $3^{rd} \text{column: } T_2$, $4^{st} \text{column: Flair}$ and image planes (top: sagittal, middle: coronal, bottom: axial). The last column is annotated image with three classes of ED, ET, NCR/NET. There exists three challenges: the imbalanced-class data, the weak boundary, and small regions of interest.}
\label{fig:datavis}
\end{figure*}

\bibliographystyle{splncs04}
\bibliography{references}